\documentclass[letterpaper,aps,prb,groupedaddress,superscriptaddress,floatfix,twocolumn,amsfonts,amssymb]{revtex4}

\usepackage{epsfig}
\usepackage{graphicx}
\usepackage{dcolumn}
\usepackage{bm}

\DeclareMathAlphabet{\mathpzc}{OT1}{pzc}{m}{it}


\newcommand{\beq}{\begin{equation}}
\newcommand{\eeq}{\end{equation}}
\newcommand{\bea}{\begin{eqnarray}}
\newcommand{\eea}{\end{eqnarray}}

\usepackage{xcolor}

\begin{document}

\title{Neural Network Solver for Small Quantum Clusters }

\author{Nicholas Walker}
\affiliation {Department of Physics and Astronomy, Louisiana State University, Baton Rouge, LA 70803, USA}

\author{Samuel Kellar}
\affiliation {Department of Physics and Astronomy, Louisiana State University, Baton Rouge, LA 70803, USA}

\author{Yi Zhang}
\affiliation {Department of Physics and Astronomy, Louisiana State University, Baton Rouge, LA 70803, USA}
\affiliation {Center for Computation \& Technology, Louisiana State University, Baton Rouge, LA 70803, USA}
\affiliation {Kavli Institute for Theoretical Sciences, University of Chinese Academy of Sciences, Beijing, 100190, China}

\author{Ka-Ming Tam}

\affiliation {Department of Physics and Astronomy, Louisiana State University, Baton Rouge, LA 70803, USA}
\affiliation {Center for Computation \& Technology, Louisiana State University, Baton Rouge, LA 70803, USA}


\date{\today}

\begin{abstract}
Machine learning approaches have recently been applied to the study of various problems in physics. Most of the studies are focused on interpreting the data generated by conventional numerical methods or an existing database. An interesting question is whether it is possible to use a machine learning approach, in particular a neural network, for solving the many-body problem. In this paper, we present a solver for interacting quantum problem for small clusters based on the neural network. We study the small quantum cluster which mimics the single impurity Anderson model. We demonstrate that the neural network based solver provides quantitatively accurate results for the spectral function as compared to the exact diagonalization method. This opens the possibility of utilizing the neural network approach as an impurity solver for other many body numerical approaches, such as dynamical mean field theory. 

\end{abstract}

\maketitle

\section{Introduction}
A single quantum impurity is the simplest possible quantum many body problem for which interaction plays a crucial role \cite{Kondo_1964,Anderson_1970}. It was invented as a model to describe diluted magnetic impurity in the otherwise non-magnetic metallic host. It has been known from the beginning that the perturbation series diverges even with an infinitesimal anti-ferromagnetic coupling strength. This became a major problem of strongly correlated systems from the the 60's \cite{Kondo_1964,Anderson_1970,Wilson_1975}. 

While the physics of a single impurity problem has been rather well studied, interest in the quantum impurity problem was revived during the 90's. This was partly due to the interest in mapping lattice models onto impurity models. \cite{Muller_Hartmann_1989a,Muller_Hartmann_1989b,Metzner_Vollhardt_1989,Bray_Moore_1980,DMFT_RMP} It has been shown that at infinite dimensions, the lattice models are equivalent to single impurity models in a mean-field as represented by the density of states of the host. This approximated mapping is known as the dynamical mean field theory. It has been further generalized to cluster impurity models to include the effects for finite dimensional systems \cite{DCA_2000,Biroli_Kotliar_2002,DCA_RMP}. 

These mappings provide a systematic tractable approximation for the lattice models and have become a major paradigm in the field of strongly correlated systems \cite{DCA_RMP}. Combined with density functional theory, they provide one of the best available methods for the study of properties of materials in which the strong interaction is important \cite{DFT+DMFT_RMP}. 

Unlike the infinite band limit which normally considered for the single impurity problem. The density of the bath, i.e. the mean-field, can be rather complicated. There is in general no analytic method for a very accurate solution. Many different methods for solving the effective impurity problem have been proposed. They can be broadly divided in the two categories: semi-analytic, and numeric. 

For the semi-analytic methods, the most widely used one is iterative perturbation theory. Its idea is to interpolate the self-energy at both the weak and strong coupling limit, together with some exact constraints, such as Luttinger theorem. \cite{Kajueter_Kotliar_1996} The other is the local moment approximation. It considers the perturbation on top of the strong coupling limit represented by the unrestricted Hatree Fock solution. \cite{Logan_Glossop_2000} 

Numerical methods can be further divided into two main classes, diagonalization based and the Monte Carlo based. Diagonalization methods usually require to discretize the host by a finite number of so-called bath sites. The Hamiltonian which includes the bath sites and one impurity site is diagonalized exactly. \cite{Caffarel_Krauth_1994} Another digonalization based method is the numerical renormalization group in which the bath sites are mapped onto a one dimensional chain of sites. The hopping amplitude decreases rapidly down the chain. The model is then diagonalized iteratively as more sites are included. \cite{Krishnamurthy_Wilkins_Wilson_1980} Density matrix renormalization group and coupled cluster theory have also been used as impurity solvers. \cite{Fernandez_Hallberg_2018,Ganahl_etal_2015,Zhu_etal_2019,Shee_Zgid_2019}. 

On the other hand, the quantum Monte Carlo method for solving impurity problems was first proposed by Hirsh and Fye \cite{Hirsch_Fye_1986}. The idea is to break up the time axis by the Trotter-Suzuki approximation. The interaction in each time segment is handled by the Hubbard-Stratonovich approximation \cite{Hirsch_1983}. The Monte Carlo method is then used to sample the Hubbard-Stratonovich fields. The idea of sampling the partition function without the Trotter-Suzuki approximation has been borrowed from the Stochastic Series Expansion in the simulation of quantum spin models to a so-called continuous time quantum Monte Carlo. \cite{Rubtsov_Savkin_Lichtenstein_2005} The method has seen a lot of development over the last decade. Notably, the expansion with respect to the strong coupling limit has been proposed and complicated coupling functions beyond simple Hubbard local density-density coupling term can now be studied. \cite{Werner_Millis_2006}

The past few years have seen tremendous development of machine learning(ML) both in terms of the algorithms and the implementation \cite{Carrasquilla_Melko_2017,Huang_Wang_2017,Wang_2016}. Many of the ML approaches in physics are designed to detect phase transitions or accelerate Monte Carlo simulations. It is a tantalizing proposal to utilize ML approaches to build a solver for quantum systems. 

A possible route to build a quantum solver based on the ML approach is to identify the feature vector (input data) and the label (output data) for the problem. Then a large pool of data are generated to train the model, specifically a neural network model. An Anderson impurity problem is a good test bed for the validity of such solver. We note that similar ideas have been explored using machine learning approaches. \cite{Arsenault_etal_2014} This paper is focused on using the kernel polynomial expansion and supervised ML, specifically, a neural network as building blocks for a quantum impurity solver. 

While it is relatively cheap to solve a single impurity problem from the above methods in the modern computational facilities, the interest in the random disorder warrants a new requirement to solve a large set of single or few impurities problems for calculating the random disorder averaging\cite{Dobrosavljevi_etal_2003,Ekuma_etal_2014,Terletska_etal_2017,Zhang_etal_2015,Dobrosavljevic_Kotliar_1998,TMDCA_review_2018,Zhang_etal_2016} . The hope is that a fast neural network based numerical solver in real frequency can expand the range of applicability of the recently developed typical medium theory for interacting strongly correlated systems, such as the Anderson-Hubbard model \cite{Ekuma_etal_2015,Ulmke_etal_1995,Semmler_etal_2011,Byczuk_etal_2005}. 


The paper is organized as follows. In the next section, we discuss mapping the continuous Green function into a finite cluster as has been done in many numerical calculations of the dynamical mean field theory. In the section III, we discuss the expansion of the spectral function in terms of the Chebyshev polynomials. In the section IV, We explain how to use the results from the section II and III as the feature vectors and labels of the neural network. In the section V, we present the spectral function calculated from the neural network approach. We conclude and discuss the future work in the last section.

\section{Representing the host by finite number of bath sites}

We first identify the input and the output data of a single impurity Anderson model. For the input data, it includes the bare density of states, the chemical potential and the Hubbard interaction of the impurity site. For a system in the thermodynamic limit, the density of states is represented by a continuous function. Representing a continuous function in the neural network presents a problem. We use the idea of exact diagonalization of the Anderson model to describe the continuous bath by a finite number of poles.\cite{Caffarel_Krauth_1994,deVega_etal_2015,Liebsch_Ishida_2011,Medvedeva_2017,Nagai_Shinaoka_2019} We first approximate the host Green function by a cluster of bath sites,
\begin{eqnarray}
G_{0}(i \omega_{n}) \approx G^{cl}_{0}(i\omega_n).
\label{map}
\end{eqnarray}

\begin{figure}[!htb]
    \centering
    \includegraphics*[height=0.18\textheight,width=0.55\textwidth, viewport=30 160 800 480,clip]{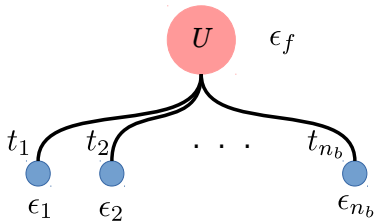}
    \caption{The cluster which represents the quantum impurity models. The red circle represents the impurity site with interaction $U$ and chemical potential $\epsilon_{f}$. The bath sites are represented by blue circles, each of them has a local energy $\epsilon_{i}$ and a hopping with the impurity site $t_{i}$}
    \label{fig:ED}
\end{figure}

In the exact diagonalization method for impurity models, the  continuum bath is discretized and represented by a finite number of so-called bath sites, see Fig.~\ref{fig:ED}. 
Assuming that there are $N_{b}$ bath sites, each bath site is characterized by a local energy ($\epsilon_{i}$) and a hopping ($t_{i}$) term with the impurity site. Two additional variables, one for the local Hubbard interaction ($U$) and the other for the chemical potential ($\epsilon_{f}$), are required to describe the impurity site. Therefore, there are in total $2+2N_{b}$ variables for representing the impurity problem. The host Green function represented in a finite cluster can be written exactly as following,

\begin{eqnarray}
G^{cl}_{0}(i\omega_{n}) = (i\omega_{n}+\epsilon_{f}-\sum_{k=1}^{N_{b}} \frac{t_k  t_{k}^{*}}{i\omega_{n}-\epsilon_{k}})^{-1}.
\end{eqnarray}

The full Hamiltonian in the discretized form is represented pictorially in Fig.~\ref{fig:ED}. It is given as 
\begin{eqnarray}
H = \sum_{i,\sigma} t_{i} (c^{\dagger}_{i,\sigma} c_{0,\sigma} + H.c.) + \sum_{i,\sigma} \epsilon_{i} c^{\dagger}_{i,\sigma}c_{i,\sigma}+ \\ \nonumber U(c^{\dagger}_{0,\uparrow}c_{0,\uparrow}-1/2) (c^{\dagger}_{0,\downarrow}c_{0,\downarrow}-1/2) - \epsilon_{f}  \sum_{\sigma} c_{0,\sigma}^{\dagger}c_{0,\sigma},
\end{eqnarray}
$c^{\dagger}_{i,\sigma}$ and $c_{i,\sigma}$ are the creation and annihilation operators for the site $i$ with spin $\sigma$ respectively. The impurity site is denoted as the $0$-th site. The sum of the bath sites are from $1$ to $N_{b}$ and the sum of the spin is for the the up and down spins for the electrons.

Parameterizing the host Green function by a finite number of variables is a standard procedure for the exact diagonalization solver for quantum impurity problems. Many different prescriptions have been investigated in details to optimize this approximation. \cite{Senchel_2010} Conceptually, practical applications of the numerical renormalization group method also require the approximated mapping onto a finite cluster chain. Unlike the exact diagonalization method, the cluster chain can be rather large, therefore much higher accuracy can be attained in general. 

The mapping onto the finite cluster to mimic the continuous bath may 
represent a nuisance. Nonetheless, this is a necessity for any diagonalization based method. These methods do not mimic the situation of continuum in the time dimension as done by continuous time Quantum Monte Carlo methods.
However the mapping presents an opportunity to naturally adapt to a machine learning approach in which a finite discretized set of variables is required.

Under the above approximation, the finite set of variables, $\{t_{i}\}, \{\epsilon_{i}\},U,\epsilon_f$
can be treated as the input feature vector for the machine learning algorithms. The next question is what is the desired output or label for the feature vectors. We will focus on the spectral function in this study. For this purpose, the next step is to represent the spectral function in a finite number of variables instead of a continuous function. We will show in the next section that the kernel polynomial method fulfills this goal. \cite{KPM_RMP}

\section{Expanding the impurity Green function by Chebyshev polynomials}

In this section, we briefly discuss the kernel polynomial method for the calculation of the spectral function of a quantum interacting model. Once the host parameters, the impurity interaction, and chemical potential are fixed, the ground state of the cluster is obtained using a diagonalization method for sparse matrix. We use the Lanczos approach in the present study \cite{Lin_1990,Lin_Gubernatis_1993}. Once the ground state is found, the spectral function can be calculated by applying the resolvent operator , $1/(\omega-H)$, to the ground state. A popular method is the continuous fraction expansion \cite{Lin_Gubernatis_1993}. The challenge is that the continuous fraction tends to be under-damped and produce spurious peaks \cite{Lin_1990,Lin_Gubernatis_1993}. A more recent method is to use the orthogonal polynomial expansion. We will argue that for the application of ML method, the polynomial expansion method tends to produce better results as we will explain later. \cite{KPM_RMP}

The zero temperature single particle retarded Green function corresponding to a generic many-body Hamiltonian is defined as 
\begin{eqnarray}
G(\omega) = \langle GS | c \frac{1}{\omega+i0^{+}-H} c^{\dagger}|GS \rangle.
\end{eqnarray}
$|GS\rangle$ is the ground state of $H$. ${c}$ and $c^{\dagger}$  are the creation and annihilation operators respectively. \cite{KPM_RMP,Wolf_etal_2014} The spectral function is given as $A(\omega)= -(1/\pi) Im(G(\omega))$. It is more convenient to directly expand the spectral function given as 
\begin{eqnarray}
A(\omega) = \langle GS | c (\omega-H) c^{\dagger} + c^{\dagger} \delta(\omega-H) c|GS \rangle.
\end{eqnarray}
Consider the Chebyshev polynomials of the first kind defined as $T_n(x) = cos(n \; arccos(x))$. Two important properties are the orthogonality and the recurrence relations. The product of two Chebyshev polynomials integrated over $x=[-1,1]$ weighted by the function $w_n = \frac{1+\delta_{n,0}}{\pi\sqrt{1-x^2}}$ is given as 
\begin{eqnarray}
\int dx w_{n}(x) T_{n}(x)T_{m}(x) = \delta_{n,m}.
\label{ortho}
\end{eqnarray} 
The recurrence relation is given as 
\begin{eqnarray}
T_{n}(x) = 2xT_{n-1}(x) - T_{n-2}(x).
\end{eqnarray}
The Chebyshev polynomials expansion method is based on the fact that
the set of Chebyshev polynomials form an orthonormal basis as defined in the Eq.~\ref{ortho}. Thus a function, $f(x)$ defined within the range of $x=[-1,1]$ can be expanded as 
\begin{eqnarray}
f(x) = \sum_{n=0}^{\infty} \mu_n(x) T_{n}(x),
\label{expansion}
\end{eqnarray}
and the expansion coefficient can be obtained by the inner product of the function $f(x)$ and the Chebyshev polynomials as follow
\begin{eqnarray}
\mu_{n} = \int_{-1}^{1}dx f(x)T_{n}(x) w_n(x).
\end{eqnarray}

Practical calculation involves truncation at a finite order. The truncation is found to be problematic, especially when the function, $f(x)$, is not smooth. For our application, the function is a spectral function of a finite size cluster, which is a linear combination of a set of delta functions. For this reason, a direct application of the above formula will not provide a smooth function. This is in analogue with the Gibbs oscillations in the Fourier expansion. The remedy is to introduce a damping factor (kernel) in each coefficient of the expansion \cite{Silver_Roder_1994,KPM_RMP,Silver_Roder_1996,Silver_Roder_1997,Alvermann_Fehske_2008}. We refer the choice of the damping factor to the review. \cite{KPM_RMP} We use the Jackson kernel given as 
\begin{eqnarray}
f(x) \approx \sum_{n=0}^{N} g_{n} \mu_n(x) T_{n}(x)
\end{eqnarray}
\begin{eqnarray}
g_{n} = \frac{(N-n+1)cos(\frac{\pi n}{N+1}) + sin(\frac{\pi n}{N+1})cot(\frac{\pi}{N+1})}{N+1}.
\end{eqnarray}

We list the steps for calculating the coefficients as follows. 

1. The input bare Green function is approximated by the bare Green function of a finite size cluster. The set of parameters $\mu, \{t_{i}\}, \{\epsilon_{i}\}$ are obtained by minimizing the difference between the left hand side and the right hand size of the Eq.~\ref{map} according to some prescriptions \cite{Caffarel_Krauth_1994,deVega_etal_2015,Liebsch_Ishida_2011,Medvedeva_2017,Nagai_Shinaoka_2019} . 

2. The ground state $(|GS\rangle)$ and the corresponding energy $(E_{GS})$ are obtained by Lanczos algorithm. 

3. The spectrum of the Hamiltonian are scaled to within the range of [-1,1] as required by the Chebyshev expansion. $H \Rightarrow (H - E_{GS}) / a$, where $a$ is a real positive constant. The units of energy are also scaled in terms of $a$. 

4. The expansion coefficients are given by the inner product between the spectral function and the Chebyshev polynomials. 
\begin{eqnarray}
\mu_n = <\alpha_0|\alpha_n>,
\end{eqnarray} 
where $|\alpha_0 \rangle = c^{\dagger} |GS>$ and $|\alpha_n \rangle = T_n(H) |\alpha_0 \rangle$. With the $|\alpha_0 \rangle$ and the $|\alpha_1\rangle=H|\alpha_0\rangle$ ready, 
all the higher order coefficients can be obtained via the recurrence relation. 
\begin{eqnarray}
|\alpha_n\rangle = 2 H | \alpha_{n-1} \rangle - |\alpha_{n-2} \rangle
\label{eq:recurrence}
\end{eqnarray}

5. The spectral function is obtained by feeding the coefficients into Eq.~\ref{expansion}.

All the coefficients can be obtained by repeated use of the Eq.~\ref{eq:recurrence} which involves matrix vector multiplication. The matrix for interacting system is usually very sparse, and the computational complexity of the matrix vector multiplication is linear with respect to the vector length, which grows as $4^{N_b+1}$ assuming no reduction by symmetries is employed. 



\section{Feature Vectors and the Labels for the machine learning}

Our strategy is to train a neural network for a large set of variables for the host, i.e., the bath sites, impurity interaction and the impurity chemical potential. The impurity solver is a function of the impurity Green function given by the bath Green function and the impurity site interaction and chemical potential, that is in total $2+2N_{b}$ variables for the input. 

The impurity Green function can be represented by $N$ coefficients of the Chebyshev polynomials expansion for the output. Using the above method the spectral function is effectively represented in terms of $N$ coefficients. It allows us to naturally employed the supervised learning method by identifying the $2+2N_{b}$ variables as the input feature vectors, and the $N$ variables as the output labels. 

While the kernel polynomial method grows exponentially with the number of sites, the end result is represented by a finite number of coefficients which presumably does not scale exponentially with the number of sites. Once the neural network is properly trained, we can use it to predict the impurity Green function without involving a calculation which scales exponentially.

\section{Results}

We generated $5000$ samples by the KPM method for randomly chosen parameters. They are drawn uniformly from the range listed as follows.
\begin{eqnarray}
t_{i,\uparrow} = t_{i,\downarrow} = [0,1.5] \\ \nonumber
\epsilon_{i,\uparrow} = \epsilon_{i,\downarrow} = [-5,5] \\ \nonumber 
U = [0,10] \\ \nonumber
\epsilon_f = [-2.5,2.5] \nonumber
\end{eqnarray}
We assume that the electron bath has a symmetric density of states. That is $t_{i} = t_{i+N_b/2}$ and $\epsilon_{i} = -\epsilon_{i+N_b/2}$ for $i = 1$ to $N_b/2$ and $N_b$ even. This further reduces the number of variables in the feature vector to $N_b+2$.

\begin{figure}[tb]
    \centering
    \includegraphics*[height=0.45\textheight,width=1.0\textwidth, viewport=00 0 800 580,clip]{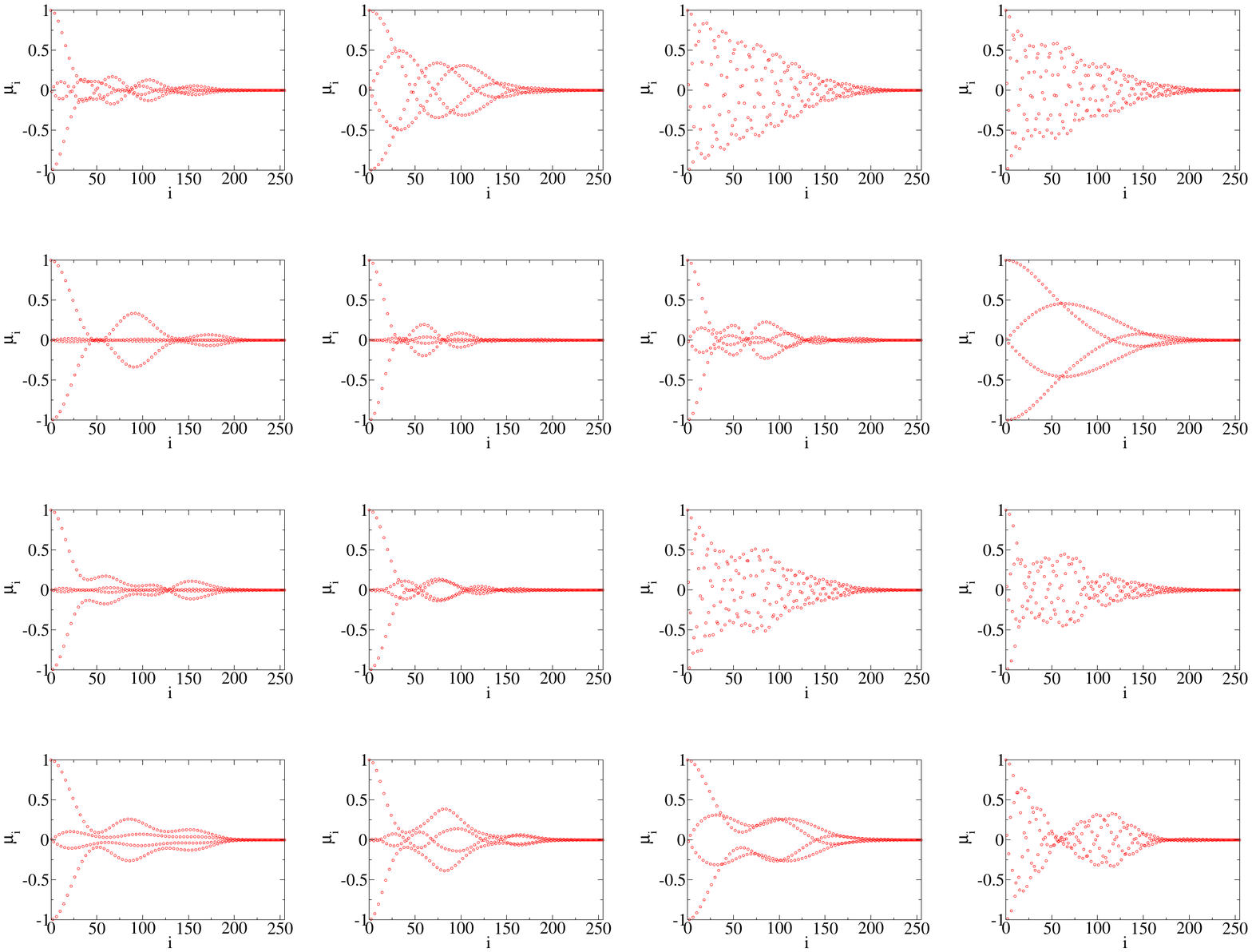}
        \includegraphics*[height=0.45\textheight,width=1.0\textwidth, viewport=00 0 800 580,clip]{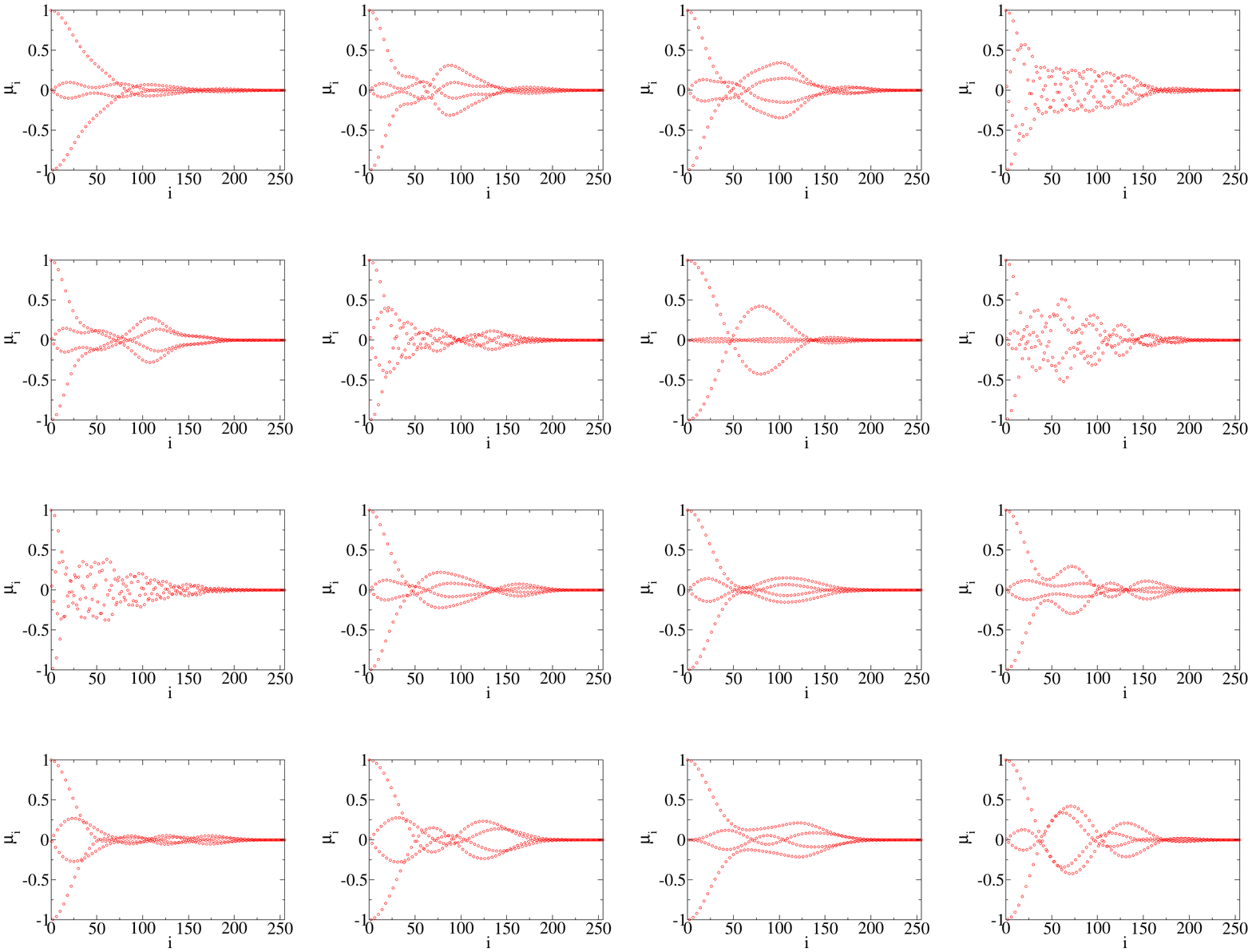}
    \caption{The coefficients of the Chebyshev polynomial expansion for 32 randomly chosen parameter sets for the finite cluster. Only the first 256 coefficients are shown, as the coefficients for the higher order terms are vanishingly small. Only the coefficients directly calculated from the kernel polynomial method (KPM)are shown here. The coefficients obtained from the neural network match very closely with the ones from the KPM and would not be visible by laying them on the same plot and thus they are omitted. We will demonstrate the quality of the coefficients in the Fig.~\ref{fig:mu_order}. The magnitude of the coefficients for the last five coefficients are smaller than $10^{-5}.$
    }
    \label{fig:mu_order}
\end{figure}

Before embarking on training the neural network, we would like to have some idea of the coefficients. We randomly pick the $32$ samples and plot the coefficients in Fig.~\ref{fig:mu_order}. There are two prominent features of the coefficients: 1. There are clear oscillations and the coefficients do not decrease monotonically; 2. For all cases shown here the coefficients essentially vanish for the orders which are around $200$ or higher. Due to these two reasons we decided to train the neural network for the coefficients from order $0$ to order $255$. 

With the above approximations, the task of solving the Anderson impurity model boils down to mapping a vector containing $N_b+2$ variables to a vector containing $N$ coefficients. For the particular case we study $N_b=6$, and $N=256$. Machine learning algorithms can thus be naturally applied to this mapping. 

We set up an independent dense neural network for each coefficient. The neural network has $14$ layers. The input layer contains $N_{in}=N_b+2$ units, and the output layer contains the expansion coefficient for one specific order. The twelve hidden layers have the number of units as follows $2N_{in}, 2N_{in}, 4N_{in}, 4N_{in}, 8N_{in}, 8N_{in},$ $8N_{in}, 8N_{in}, 4N_{in}, 4N_{in}, 2N_{in}, 2N_{in}$. 

As we consider in total of $256$ orders, we have $256$ independent neural networks. Considering the coefficients at different orders separately may lose some information contained in the correlations between different orders. While it is possible to predict a few coefficients by one neural network, we do not get a good prediction for using a single neural network to predict all $256$ coefficients without an elaborated fine tuning. Therefore, instead of searching for a optimal number of coefficients for one neural network, we consider each coefficient independently. 

\begin{figure}[!htb]
    \centering
    \includegraphics*[height=0.45\textheight,width=1\textwidth, viewport=00 0 800 580,clip]{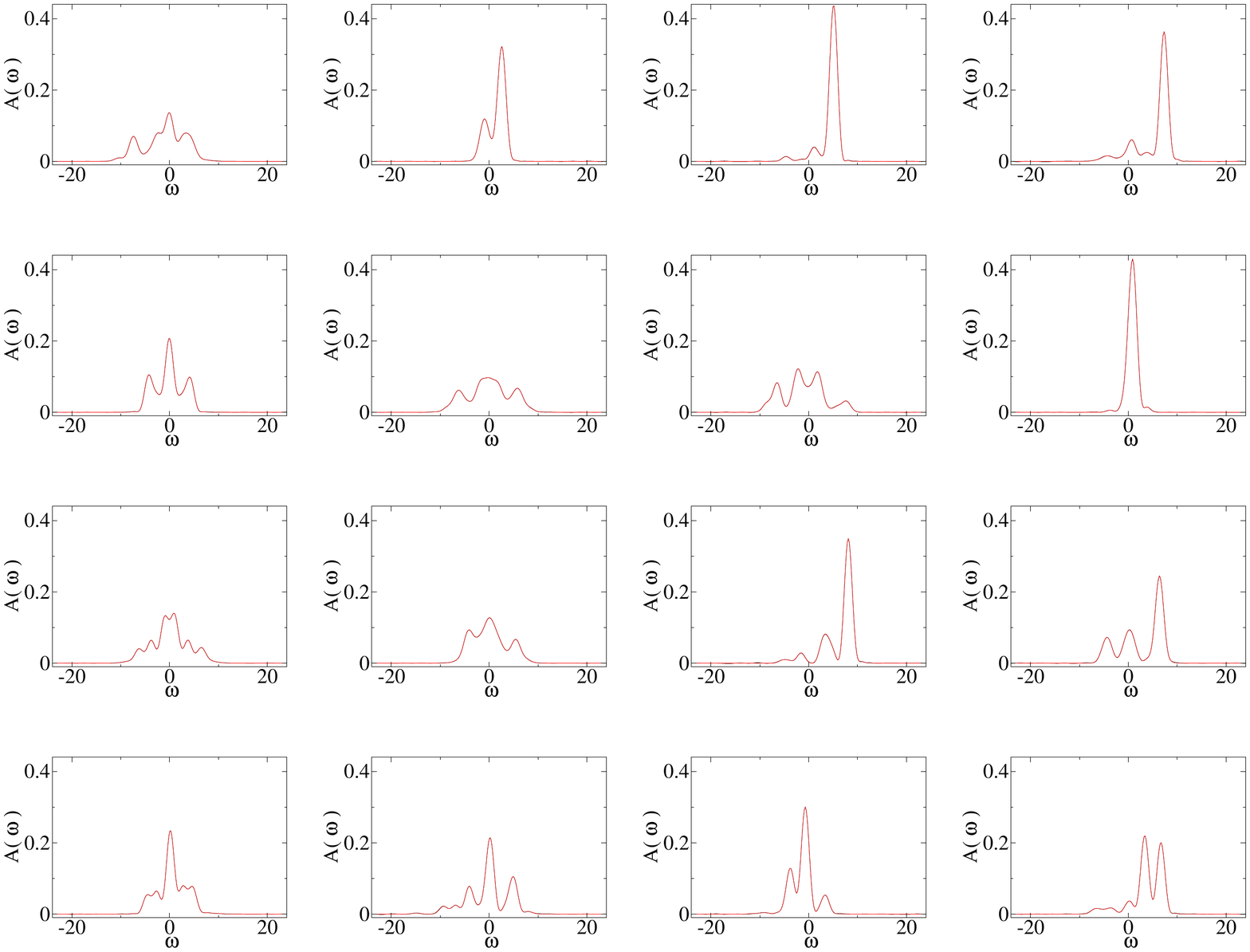}
        \includegraphics*[height=0.45\textheight,width=1\textwidth, viewport=00 0 800 580,clip]{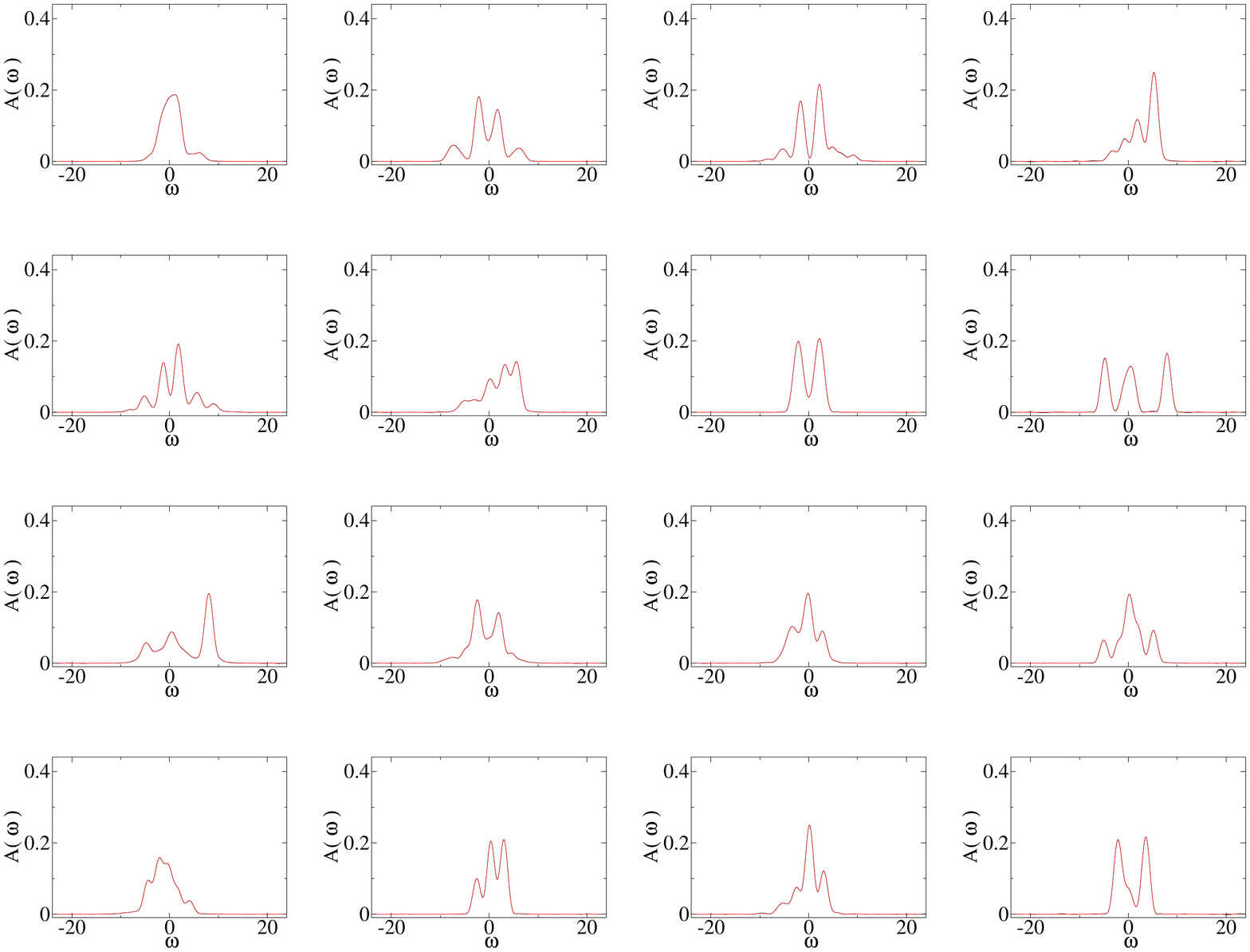}
    \caption{The spectral function, $A(\omega)$ is plotted for 32 randomly chosen parameter sets for the finite cluster. The figures correspond to the coefficient as shown in the Fig.~\ref{fig:mu_order}. Both the results from the KPM and from the neural network are shown. They match each other very closely, and visually overlap on top of each others. A closer inspection reveals that there are slight oscillations in the spectral function when the weights are very small. This may due to the in-exact cancellations of different orders from the coefficients generated by the neural network method. In general these oscillations are rather small and only appear when the spectral weight drops to near zero. 
        }
    \label{fig:spectral}
\end{figure}

\begin{figure}[!htb]
    \centering
    \includegraphics*[height=0.45\textheight,width=1\textwidth, viewport=00 0 800 580,clip]{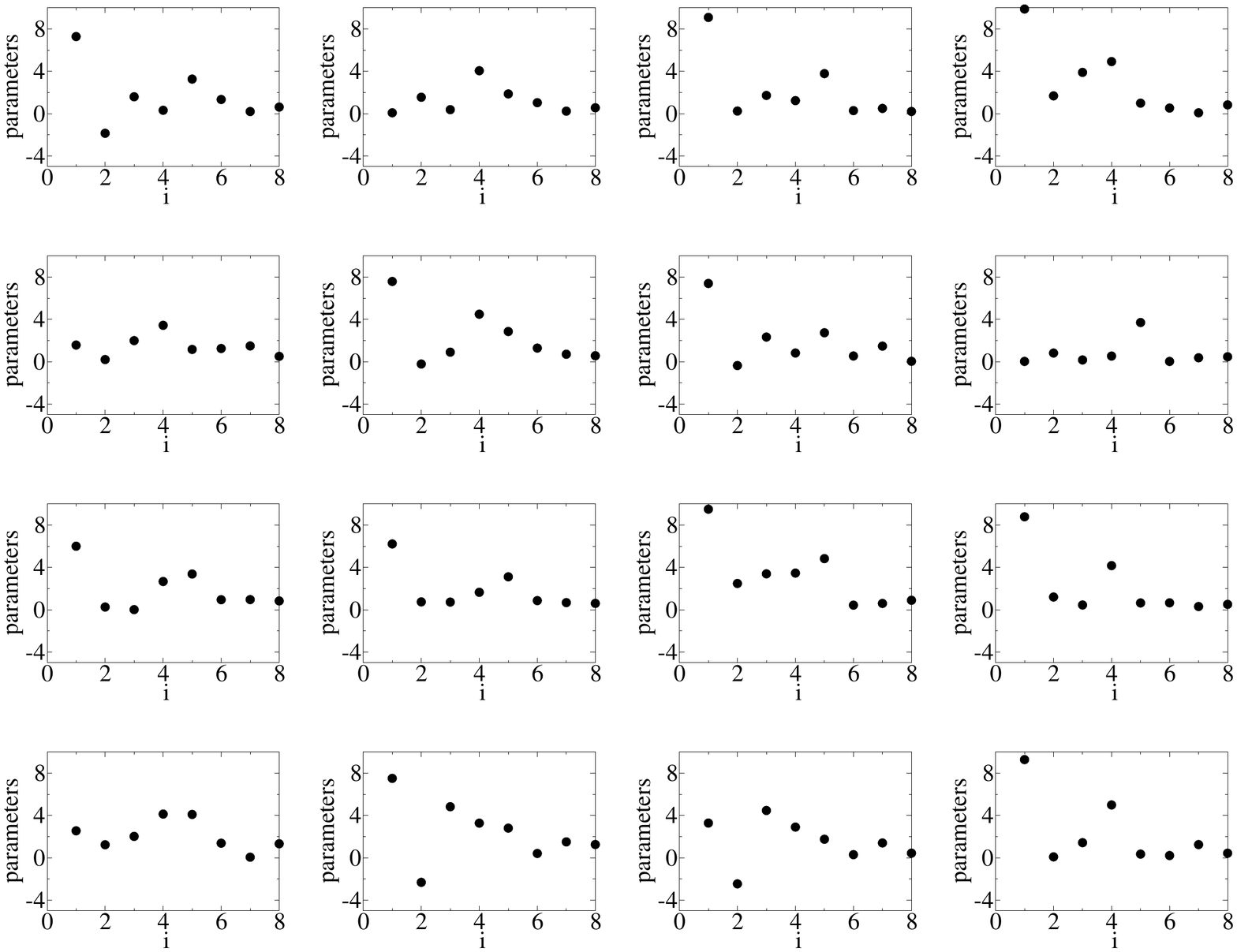}
        \includegraphics*[height=0.45\textheight,width=1\textwidth, viewport=00 0 800 580,clip]{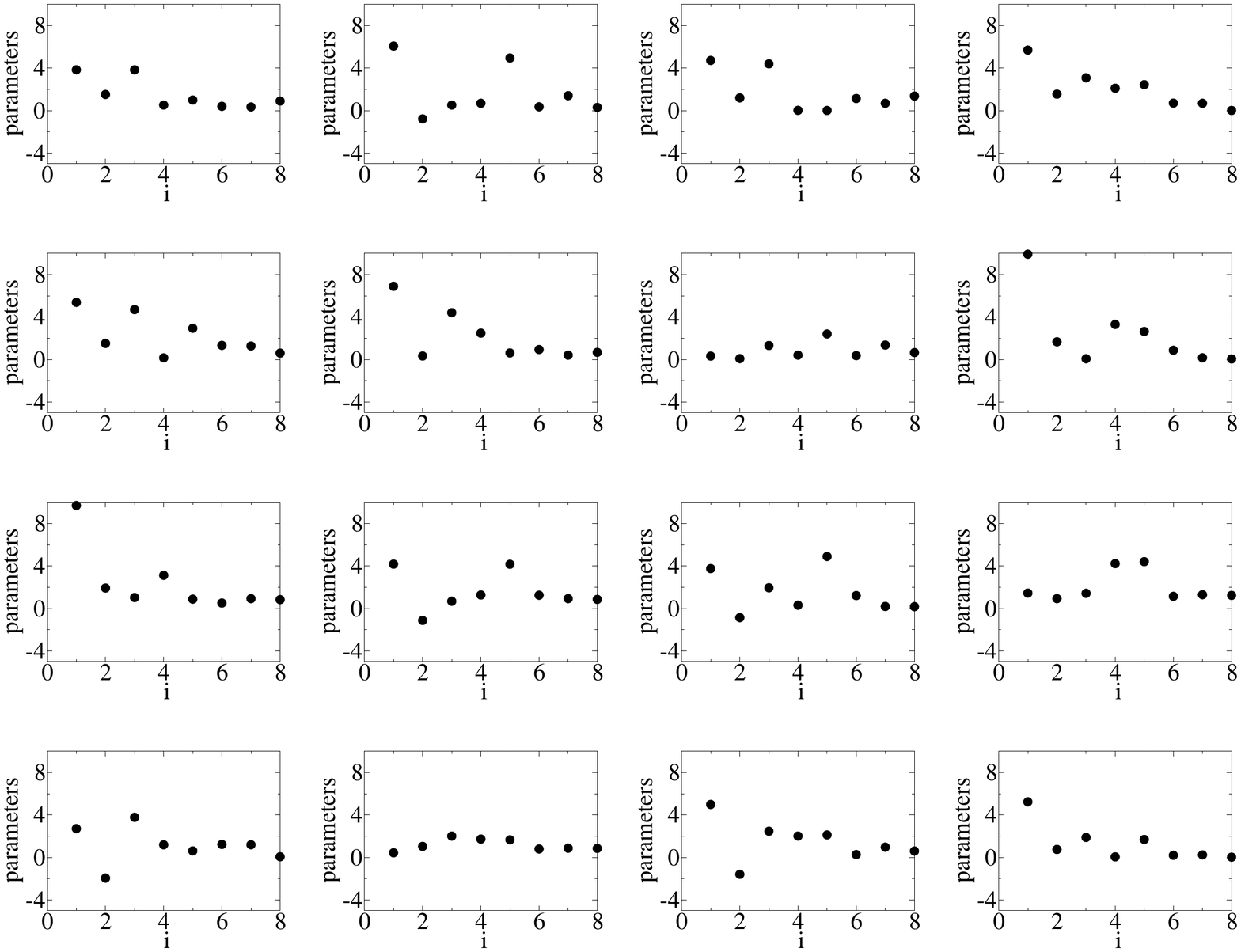}
    \caption{The input parameters of the $32$ density of states plotted in Fig.~\ref{fig:spectral}. $i=1$ corresponds to $U$, $i=2$ corresponds to $\epsilon_{f}$, $i=3,4,5$ correspond to $\epsilon_1,\epsilon_2,\epsilon_3$ and $i=6,7,8$ correspond to $t_1,t_2,t_3$.
    }
    \label{fig:param}
\end{figure}

We show the spectral function in the Fig.~\ref{fig:spectral}, they are from the same $32$ samples as that in the Fig.~\ref{fig:mu_order}. Both the results from the direct numerical calculation based on the Lanczos method and recurrence relation and those from the neural network prediction are plotted. They basically overlap with each other. There is a slight difference for the range of energy where the spectral function is nearly zero. This is perhaps due to the incomplete cancellation among the expansion terms at different orders due to the errors from the neural networks. An improvement may be attainable if we consider the correlations of the coefficients for different orders. The input parameters of each of the $32$ samples are plotted in Fig.~\ref{fig:param}.

\begin{figure}[!htb]
    \centering
    \includegraphics*[height=0.45\textheight,width=1\textwidth, viewport=00 0 800 580,clip]{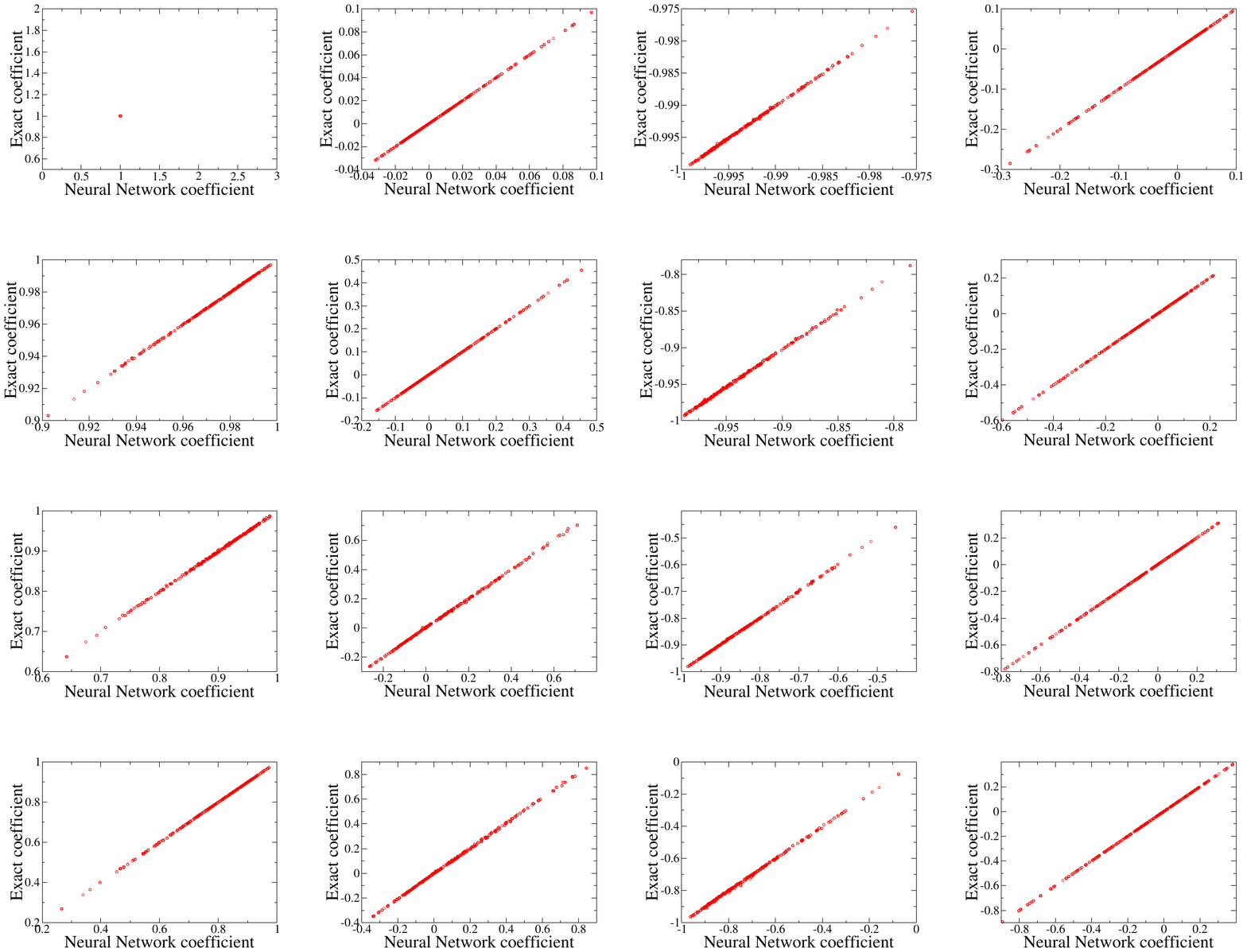}
        \includegraphics*[height=0.45\textheight,width=1\textwidth, viewport=00 0 800 580,clip]{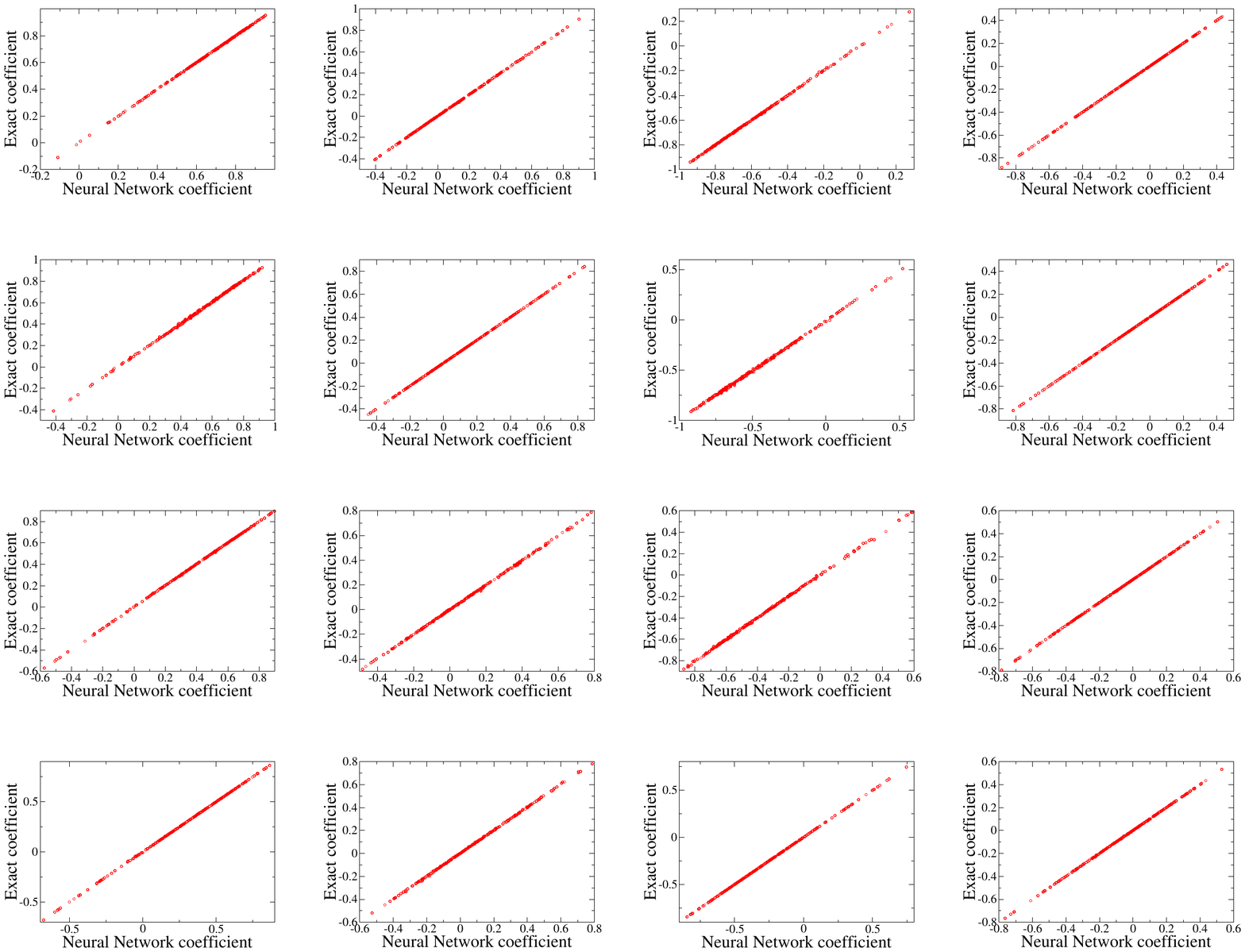}
    \caption{Comparisons of the first 32 coefficients as computed by the KPM and the neural network method. 1000 samples are plotted in each figure. The figures are ordered from left to right and top to bottom from order the 0-th to the order 31-th. 
    }
    \label{fig:mu}
\end{figure}

Evidence of the capability of the neural network approach can be seen in Fig.~\ref{fig:mu}, we plot the comparisons of the first $32$ expansion coefficients obtained by the direct numerical calculation and the neural network prediction. $1000$ samples are considered, we find that two methods give very close results. 
With the $1000$ samples we considered, all exhibit a linear trend. This clearly shows that a neural network is capable of providing a good prediction. There were no exceptional outlier among the $1000$ samples we tested.

\section{Conclusion}
We demonstrate that the supervised machine learning approach, specifically the neural network method, can be utilized as a solver for small quantum interacting clusters. This could be potentially useful for the 
statistical DMFT or typical medium theory for which a large number of
impurity problems have to be solved for disorder averaging. \cite{Dobrosavljevi_etal_2003,Ekuma_etal_2014,Terletska_etal_2017,Zhang_etal_2015,Dobrosavljevic_Kotliar_1998,Ekuma_etal_2015} The main strategy is to devise a finite number of variables as the feature vector and the label for the supervised machine learning method. In line with the exact diagonalization method for the single impurity Anderson model, the feature vector is represented by the hopping and the local energy of the lattice model. The output, spectral function, is represented in terms of Chebyshev polynomials with the damping kernel. The label is then represented by the coefficients of the expansion. By comparing the coefficients directly calculated by the Lanczos method and the recurrence relation and that by the neural network, we find the agreement between the results from these two methods is very good. Notably, among the $1000$ samples being tested, there is no exceptional outlier. They all have good agreement with that from the direct numerical method. 

For a simple impurity problem, the present method may not have an obvious benefit, as a rather large pool of samples have to be generated for training at the first place. The situation is completely different for the study of disorder models, such as those being studied by the typical medium theory, where the present method has a clear advantage. Once the neural network is trained the calculations is computationally cheap. For systems in which disorder averaging is required, this method can beat most if not all numerical methods in term of efficiency. Moreover the present approach is rather easy to be generalized for more complicated models, such as a few impurities model required in the dynamical cluster approximation. In addition, the matrix product basis has been proposed for the kernel polynomial expansion, this method can be easily adapted for it. \cite{Wolf_etal_2014,Wolf_etal_2015}

The ideas presented in this paper is rather generic. They can be generalized for the solutions from different solvers. For example, it can be adapted to the solutions from QMC as long as the solutions can be represented in some kind of series expansion \cite{Huang_2016,Boehnke_etal_2011} and it can also be adapted for the coefficients from the coupled cluster theory \cite{Zhu_etal_2019,Shee_Zgid_2019}.

\section{Acknowledgement}
We thank Mark Jarrell for his comments and suggestions for this project. This work is funded by the NSF Materials Theory grant DMR1728457. This work used the high performance computational resources provided by the Louisiana Optical Network Initiative (http://www.loni.org) and HPC@LSU computing. Additional support (NW and KMT) was provided by the NSFEPSCoR CIMM project under award OIA-154107. Additional support (YZ) at Louisiana State University was provided by the U.S. Department of Energy, Office of Science, Office of Basic Energy Sciences under Award No. DE-SC0017861. 
\bibliography{refs}

\onecolumngrid

\twocolumngrid

\onecolumngrid

\twocolumngrid

\onecolumngrid

\twocolumngrid

\onecolumngrid

\twocolumngrid

\end{document}